# Nanosculpting lateral weak link junctions in superconducting Fe(Te,Se)/Bi$_2$Te$_3$ with focused Si$^{++}$ ions and implications on vortex pinning

*Debarghya Mallick[1,\*], Sujoy Ghosh[2], An Hsi Chen[1], Qiangsheng Lu[1], Liam Collins[2], Sangsoo Kim[1], Gyula Eres[1], Ivan Kravchenko[2], Stephen Jesse[2], Steven J. Randolph[2], Scott T. Retterer[2], Matthew Brahlek[1], Robert G. Moore[1,\*]*

[1]Materials Science and Technology Division
Oak Ridge National Laboratory
Oak Ridge, TN, 37831, USA X

[2]Center for Nanophase Materials Sciences
Oak Ridge National Laboratory
Oak Ridge, TN 37831, USA

[\*]Correspondence
Email: mallickd@ornl.gov
Email: moorerg@ornl.gov

**Abstract:**
Superconductor–normal–superconductor (SC-N-SC) weak links enable Cooper-pair tunnelling and serve as Josephson junctions (JJs) used in modern superconducting qubits. Conventional JJs rely on vertically stacked Al-AlO$_x$-Al trilayers that are difficult to fabricate and are sensitive to ambient exposure. Here, we demonstrate an all-in-plane alternative by "nanosculpting" ~100 nm-wide channels into thin films of FeTe$_{0.75}$Se$_{0.25}$/Bi$_2$Te$_3$ (FTS/BT), a candidate topological superconductor, with a Si$^{++}$ focused ion beam (FIB). Systematic irradiation shows that increasing the ion dose, while keeping the beam energy constant, progressively suppresses both the critical temperature ($T_c$) and critical current ($I_c$), confirming the creation of a controllable weak link even though a Fraunhofer interference pattern is not observed. Kelvin probe force microscopy, atomic force microscopy, and scanning electron microscopy corroborate the structural and electronic modification of the irradiated region. $I_c$ ($B$) measurements reveal a slower field-induced decay of $I_c$ at higher doses, indicating that irradiation-induced defects act as vortex-pinning centers that mitigate vortex motion and associated dissipation. By tuning beam energy and dose, the process shifts from an SC-N-SC regime toward a superconductor–insulator–superconductor (SC-I-SC) geometry, offering a simple scalable pathway to JJ fabrication. These results establish FIB patterning as a versatile platform for engineering robust, scalable fault-tolerant qubits.

## 1. Introduction

The discovery of iron-based superconductivity in LaFeAsO in 2008 sparked significant interest in the scientific community [1,2]. Since then, numerous iron-based superconductors (FeSCs), including pnictides and chalcogenides, have been identified with $T_c$ reaching up to 100 K [3-5]. It remains puzzling that superconductivity can exist in materials containing iron, which is typically magnetic. The pairing mechanism in FeSCs remains a topic of debate [6,7]. Theoretical studies suggest that superconducting pairing in FeSCs may originate from the interplay of magnetism, which is intriguing given that magnetism and superconductivity are traditionally considered antagonistic phenomena.

FeSe has garnered significant attention due to the notable enhancement of $T_c$ at the monolayer limit and recent observations consistent with topological superconductivity in $FeTe_{0.55}Se_{0.45}$ [8-13]. These observations have fostered interest in the $FeTe_{0.55}Se_{0.45}$ (FTS) system as a platform for hosting Majorana modes with non-Abelian statistics for quantum computing applications [14]. However, $FeTe_{0.55}Se_{0.45}$ suffers from a major challenge: the inhomogeneity in the Te/Se ratio, which leads to local fluctuations in their topological and superconducting properties [15]. One promising approach to address this issue is to create a heterojunction between a superconducting film and a topological insulator, leveraging proximity-induced topological superconductivity at an epitaxial interface [16]. This configuration potentially allows a more uniform superconducting layer in Fe(Te,Se) to imprint superconductivity on the interfacial topological spin textures via the proximity effect, offering a more homogeneous platform to realize topological superconductivity [17-24]. With the heterostructure approach comes the need to fabricate quantum devices capable of Majorana fusion and braiding operations. This approach also requires controlling the location and mobility of superconducting vortices in an applied magnetic field, where the Majorana modes exist.

Enhancing vortex pinning has been an important area of research for the past three decades, after the discovery of high $T_c$ superconductors, due to the high critical field and temperature of these materials [25-27]. However, one of the main challenges has been the high cost of manufacturing superconducting wires from ceramic high $T_c$ superconductors [28]. Extensive efforts have been made to improve the critical current in high $T_c$ superconductors by increasing the pinning of the vortex through ion irradiation with various species. Since the discovery of FeSCs, these materials have attracted attention due to their higher upper critical field ($B_{c2}$) and moderately high $T_c$. Although not as high as cuprates, they surpass conventional superconductors, making them strong candidates for industrial applications and eventual commercialization. Hence, understanding and controlling vortices in FeSCs has wide-ranging implications, from quantum computing to industrial applications. Recent efforts to increase the critical current ($I_c$) in FeSCs have focused mainly on macroscopic samples subjected to ion irradiation, with limited exploration of dose dependence [29,30,30-34]. Furthermore, reports on irradiation using protons, gamma rays, and other energetic particles have not used focused beams, which can locally modify the superconductor on the nanoscale [30,31,33-37]. Little work has been done on the effects of focused ion beam (FIB) irradiation, especially with $Si^{++}$ ions, on superconductor devices. Investigating the dose dependence of focused ion irradiation on FeSCs could yield valuable insights for both manipulating magnetic vortices and fabricating functional quantum devices.

Josephson junctions (JJs) play a crucial role in superconducting circuits, contributing significantly to the two-level system of qubits. There are various methods to fabricate JJs, with the vertical Al-AlO$_x$-Al structure being the most common [38]. A more recent approach involves using a focused ion beam (FIB) to irradiate superconductors [39-41]. In this process, the energetic ion beam interacts with the superconducting lattice at the nanoscale, perturbing the material and forming a junction. For this junction to function as a JJ, supercurrent must tunnel through the potential barrier created by the irradiation. If the barrier is too weak, Cooper pairs can tunnel freely with no significant effect. Conversely, an excessively high energy beam or excessive energy deposition can create a barrier that is too strong, preventing supercurrent from tunneling and resulting in an absence of diffraction effects, such as the Fraunhofer pattern. To achieve optimal junction behavior, it is essential to finely tune parameters such as the ion beam species, energy, and dose. Consequently, optimizing the FIB dose is of paramount importance to regulate the barrier height and to investigate the effects of focused ion beam irradiation on topological superconductor (TSC) thin films. Previously, several studies have explored the effects of irradiation on Fe(Te,Se) (FTS) superconductors [36, 37, 55]. However, in those investigations, the entire samples were subjected to irradiation. To date, there have been no reports focused on the technique of nano-sculpting within iron-based superconductors to create a weak link, thereby establishing a superconducting-non-superconducting-superconducting architecture.

In this study, we examine the effect of an energetic $Si^{++}$ ion beam, using a FIB system, on a thin film of the topological superconductor candidate $FeTe_{0.75}Se_{0.25}/Bi_2Te_3$ (FTS/BT) by varying the dose of the ion beam over a nanoscale region, starting at about 150 nm while keeping the beam energy constant. We introduce defects in a controlled manner by adjusting the ion beam dose in a localized region of the heterostructure, thereby affecting superconductivity within that nano-sized region, enabling to achieve the superconductor-FIB irradiated region-superconductor lateral structure. This method offers a pathway to tune the dose and energy of the $Si^{++}$ beam to create a weak link for Cooper pair tunneling [41,42], potentially enabling the fabrication of fault-tolerant topological qubits. Specifically, we study superconducting parameters such as $T_c$ and $I_c$ by applying the FIB in a line scan at different doses on a micrometer-scale bridge device. We also unveil the effect of vortex-pinning by demonstrating a slower rate collapse in $I_c$ with $B$ owing to the reduction in the vortex creep motion. Understanding the effects of $Si^{++}$ irradiation on a topological superconductor/topological insulator heterojunction device will pave the way for planar Josephson junctions, ultimately contributing to scalable fault-tolerant qubit fabrication as well as further our understanding of the underlying superconducting properties [43,44].

## 2. Results and Discussion

An epitaxial $FeTe_{0.75}Se_{0.25}$ (FTS) thin film (12 nm) was grown on a $Bi_2Te_3$ (BT) layer (50 nm) on an $Al_2O_3(0001)$ substrate using molecular beam epitaxy (MBE), as described in Ref.[23]. A schematic of the heterostructure stack is shown in **Figure** 1a. Structural characterization via x-ray diffraction (XRD) and atomic force microscopy (AFM) can also be found in Ref.[23]. $Bi_2Te_3$ has a triangular surface, whereas Fe (Te,Se) has a square surface. Despite this symmetry mismatch, these two materials are found to epitaxially compatible. This is largely driven by the nearly commensurate lattice spacing of the $Bi_2Te_3$ orthogonal to the <210> ($4.38 \times \frac{\sqrt{3}}{2} = 3.80$ Å) direction to that of Fe (Te, Se) along the <100> (3.81Å for x=0.25), which has been called hybrid epitaxy [21]. Using standard photolithography followed by $Cl_2$-gas based reactive ion etching, the entire 10 mm × 10 mm heterostructure chip was patterned into an array of 16 Hall bar devices, each with a 4 $\mu$m channel width. Subsequently, each device was irradiated with a focused $Si^{++}$ ion beam (35 kV) of diameter 5–10 nm, as depicted in Figure 1b. The ion doses varied from 0 pC/$\mu$m (i.e., no irradiation) to 200 pC/$\mu$m. If the ion used is too light, such as $He^{2+}$, it may not significantly affect the superconducting properties. Conversely, using a commonly employed heavy ion like $Ga^{2+}$ can result in substantial damage or even sputtering of the material, even at lower doses.

Therefore, we selected Si²⁺ ions for this initial study, as they strike an optimal balance—being neither too light nor excessively heavy—thus allowing for effective modification of the material while minimizing unwanted damage. The scanning electron microscope (SEM) image in Figure 1c illustrates the four electrical transport probes (current and voltage contacts). The inset of Figure 1c highlights the Si$^{++}$-irradiated region (150 nm wide for this device), which is faintly visible within the Hall bar channel (green box in the main figure). Upon entering the sample, the energetic Si$^{++}$ ions undergo multiple collisions and scattering events, ultimately settling in different layers of the heterostructure. "Stopping and Range of Ions in Matter (SRIM)" simulations, shown in Figure 1d, depict the trajectories of implanted ions within the heterostructure [45]. The inset shows the ion distribution profile, indicating that the highest density of ions is localized within the FTS layer, directly influencing its superconducting properties. The vertical lines provide a schematic representation of the layered structure. Although the SRIM simulations were performed assuming an amorphous sample, this approximation does not significantly alter the qualitative picture of the ion trajectories. Figure 1e displays an atomic force microscopy (AFM) image (left) and a Kelvin probe force microscopy (KPFM) map (right) recorded from the same device region after irradiation (with a dose of 10 pC/μm). The AFM image resolves the surface topography of the ion-milled line and surrounding channel, whereas the KPFM map visualizes the spatial variation of the local work function. The irradiated line exhibits a systematic work function increase of ∼ 10–15 mV relative to the pristine area, as highlighted in the KPFM panel. Because absolute work function values may be subject to calibration offsets, the primary conclusion is that there is a relative shift between irradiated and unirradiated regions. The microscopic origin of this increase is uncertain, as implanted Si$^{++}$ ions may introduce local charge doping, while irradiation-induced lattice defects or strain may also contribute. Regardless of the mechanism, such local work function modulation must be considered when designing and fabricating devices [46].

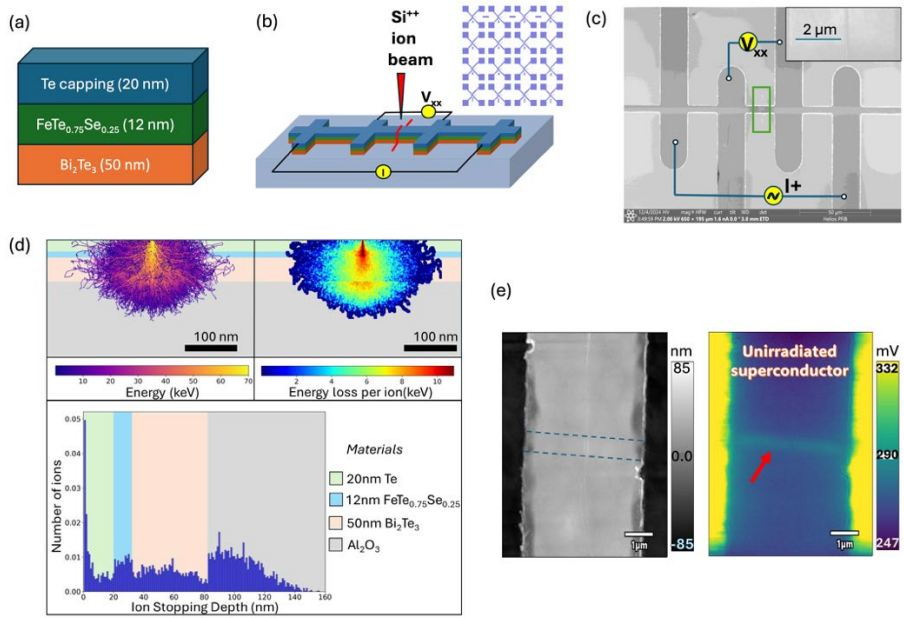

**Figure 1.** Device image, Schematics and Ion trajectory simulation: (a) A cross-sectional schematic of the heterostructure, with layer thicknesses specified in brackets. (b) A schematic illustrating the focused ion beam of Si$^{++}$, irradiating a device fabricated from the FeTe$_{0.75}$Se$_{0.25}$/Bi$_2$Te$_3$ heterostructure. The inset shows an array of 16 such devices on a single 10 mm × 10 mm chip, each subjected to varying ion beam doses. (c) A scanning electron microscopy (SEM) image highlights the central region of each device, featuring four contact pads and a ~ 5 µm-wide channel. The irradiated region can be observed faintly inside the green box and is further magnified in the inset. (d) SRIM simulation of 5,000 Si$^{++}$ 70 keV ions at into the heterostructure. The ion trajectories are shown in the top left image, where the ion energy at a given collision is given by the color scaling. The average energy loss per ion is shown in top right, where distinct discontinuities arise at the interfaces. The highest energy losses occur in the Te cap and the FTS layers. The normalized distribution of ions at the end of a trajectory is shown in the bottom panel. The peak at the surface is indicative of backscattered ions. The FTS region of the sample shows increased stopping of ions compared to the upper Te layer. Once ions traverse the Bi$_2$Te$_3$ layer, they are then effectively stopped in the sapphire substrate. (e) AFM (left) and KPFM (right) images show the ~ 5 µm wide channel. The irradiated region (with a dose of 10 pC/µm) towards the middle is marked by the dotted line in the left image and indicated by the red arrow in the right image.

If the sample is continuously bombarded with ions at progressively higher doses, material begins to sputter from the surface. Beyond a certain threshold dose, this sputtering process disrupts and removes the material entirely, resulting in a discontinuous channel. This effect is illustrated in **Figure** 2. We irradiated the sample following the same geometry shown in Figure 1a, starting from 0 pC/µm and incrementally increasing the dose, while simultaneously recording the current *in situ* (micromanipulator probes are shown as an inset of Figure 2) under a fixed 1 V bias at room temperature. Initially, the channel exhibits a resistance of approximately 2 kΩ in the absence of ion irradiation. As the dose increases, the current remains largely unchanged, indicating minimal damage at lower doses. However, around a 300 pC/µm dose, the current begins to drop abruptly, marking the onset of severe structural disruption. By ~325 pC/µm, the current falls to zero, confirming that the channel is completely removed due to extensive sputtering. The consistency of this behavior across multiple devices underscores its reproducibility. As such, for further electrical transport studies, we kept the dose well below the critical sputtering threshold (300 pC/µm).

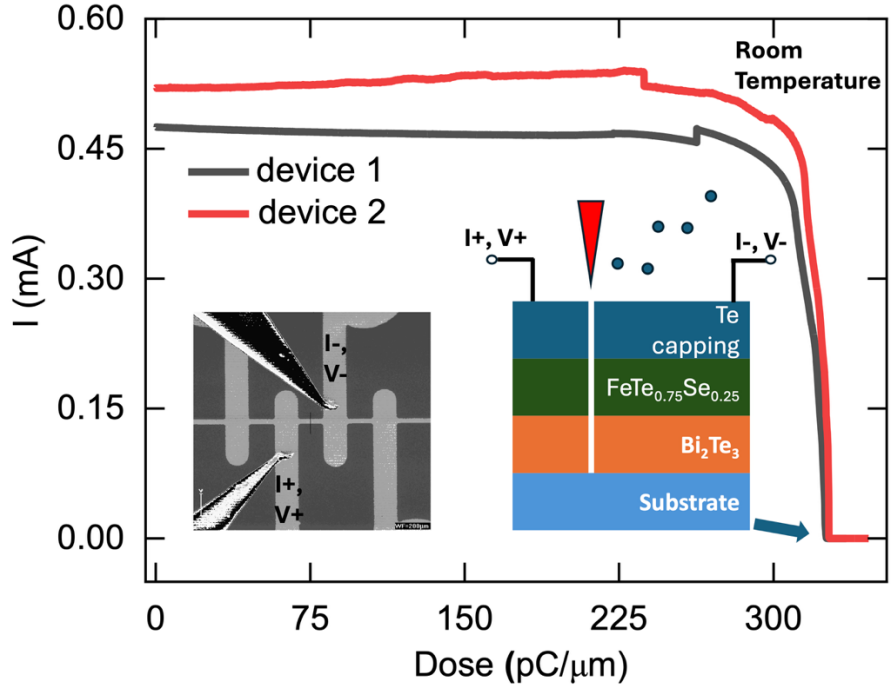

**Figure 2.** Real Time Current vs. Dose for a Device Under a Fixed Bias. The figure presents the current flowing through the device channel under a constant voltage bias of 1 V while a focused Si$^{++}$ ion beam is irradiated on a localized region of the channel. The dose increases linearly with time from 0 pC/µm to the point where the thin film becomes completely discontinuous (approximately 325 pC/µm). The current remains nearly constant

up to a dose of around 250 pC/μm, after which it sharply drops to zero. This indicates that significant material removal due to ion beam sputtering begins at a dose of approximately 300 pC/μm, eventually severing the channel at 325 pC/μm. The near constancy of current up to 250 pC/μm suggests that physical damage from sputtering is negligible up to this point. Instead, the ion beam likely induces lattice distortions and defects, subtly affecting the superconducting properties before substantial material removal occurs. The left inset displays an SEM image of the device during the current vs. dose measurement, showing the two probes used for biasing and current measurement. The right inset illustrates the moment when the channel becomes fully removed, as indicated by the blue arrow. The red narrow triangle represents the focused ion beam, while the blue dots depict sputtered material due to ion bombardment.

Resistance-versus-temperature (*R-T*) measurements were then performed for devices irradiated with varying ion doses (**Figure** 3a). The left panel of Figure 3a shows that $T_c$ of a pristine device (i.e., without ion irradiation) is approximately 9.5 K. As the ion irradiation dose increases, $T_c$ progressively decreases, as shown in Figure 3b. At the highest applied dose of 200 pC/μm, $T_c$ drops to around 8 K. This decrease in $T_c$ is expected, as energetic ions passing through the layers introduce lattice damage, modifying local electrical properties and weakening superconductivity in the irradiated regions [47]. It is important to note that the $T_c$ of a superconducting layer and that of a superconducting-non superconducting- superconducting (S-S'-S) structure should not differ significantly, as long as the superconductivity in the non-superconducting layer (S') is weakly suppressed so the cooper pair can tunnel through the barrier. The minimal change in Tc in Figure 3b (~ 1.5 K) is consistent with our approach towards establishing a weak link.

**Figure 3.** Resistance vs. Temperature Behavior of the Junction. (a) Resistance vs. temperature plots for different doses of focused $Si^{++}$ ion beam irradiation. The data show that superconductivity is weakened at sufficiently high doses. (b) The superconducting transition temperature ($T_c$) extracted from the curves in (a), is plotted as a function of ion dose. The decreasing trend of $T_c$ with increasing dose is expected, as the energetic $Si^{++}$ ion bombardment weakens superconductivity in the irradiated region. The blue dotted line indicates the range of $T_c$ values, spanning from the pristine (non-irradiated) device to the device subjected to the highest ion dose.

Next, we investigated the current–voltage (*I-V*) characteristics of a pristine superconducting heterostructure at T = 2.2 K, examining their evolution under varying magnetic fields and temperatures. **Figure** 4a shows the *I–V* curves at *T* = 2.2 K for perpendicular magnetic fields ranging from 0 T to 8.5 T. At *B* = 0 T, the device is fully superconducting, exhibiting a characteristic *I-V* response: a nearly flat region near *I* = 0 corresponding to the superconducting state, followed by an Ohmic region at higher current once superconductivity is suppressed. The critical current ($I_c$) is defined as the current at which superconductivity vanishes. Figure 4c plots $I_c$ for different magnetic fields. As *B* increases, vortices penetrate the sample and progressively suppress superconductivity, causing $I_c$ to decrease [34]. The critical current follows a power law dependence on $B$, $I_c \sim B^{-\gamma}$, where $\gamma$ characterizes vortex-vortex interactions [37,48]. When vortices are generated due to magnetic flux penetration, they begin to exhibit creep motion. The extent of this vortex creep enhances the interaction between neighboring vortices, resulting in a higher value of $\gamma$. By comparing the $\gamma$ values, one can qualitatively assess the degree of vortex creep in different superconductors. Figure 4b shows the temperature dependence of $I_c$ in a pristine device: as the temperature increases, $I_c$ decreases, and beyond $T_c \sim 10$ K the device becomes normal with a linear (Ohmic) $I$-$V$ response.

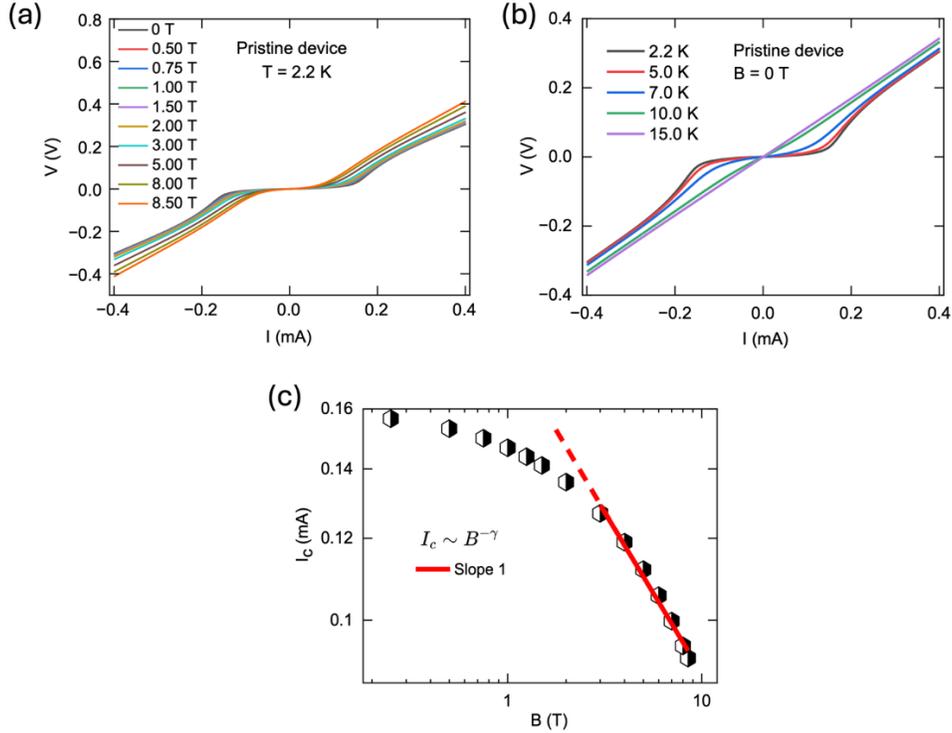

**Figure 4.** Current-Voltage Characteristics of a Pristine Device. (a) Current-voltage characteristics of the pristine device without any irradiation, measured at different perpendicular magnetic fields at the base temperature, $T =$ 2.2 K. (b) Temperature dependence of the $I$-$V$ characteristics for the same pristine device in the absence of an external magnetic field. In both cases—variation with magnetic field and temperature—the $I$-$V$ curves gradually transition toward the Ohmic regime, indicating the suppression of superconductivity. (c) The critical current ($I_c$), extracted from the curves in (a), is plotted as a function of the magnetic field on a log-log scale. A fitting is performed to determine the slope, which qualitatively characterize the vortex dynamics induced by the applied magnetic field. The greater the vortex creep motion induced by the applied magnetic field, the more rapidly $I_c$ collapses, resulting in a steeper slope.

Having established the dose-dependent sputtering threshold of our sample, we selected specific ion doses to irradiate devices at $T = 2.2$ K to investigate how the critical current evolves with increasing dose. The corresponding $I$-$V$ characteristics, shown in **Figure** 5a, span ion doses from 0 to 200 pC/$\mu$m. For this study, the doses were kept well below 300 pC/$\mu$m to avoid substantial sputtering. As seen in Figure 5a, increasing the ion dose gradually influences the superconducting state, apparent from the $I$-$V$ curves. Nevertheless, even at the highest dose of 200 pC/$\mu$m, superconductivity is largely preserved, with only a minor resistive component—observed as a small non-zero gradient near zero magnetic field. We hypothesize as follows: up to ~250 pC/$\mu$m, most $Si^{++}$ ions embed within the lattice without causing significant material loss. Beyond this threshold, ion bombardment commences sputtering away the material, severely disrupting the channel. The extracted $I_c$ values from each $I$–$V$ curve are plotted as a function of dose in Figure 5b. Interestingly, the critical current (Ic) does not exhibit a monotonic relationship with the irradiation dose. As illustrated in Figure 5b, Ic decreases sharply up to approximately 60 pC/μm, after which it levels off and remains relatively constant through to the highest dose of 200 pC/μm. This behavior suggests that superconductivity is highly sensitive to low-dose irradiation compared to higher doses.

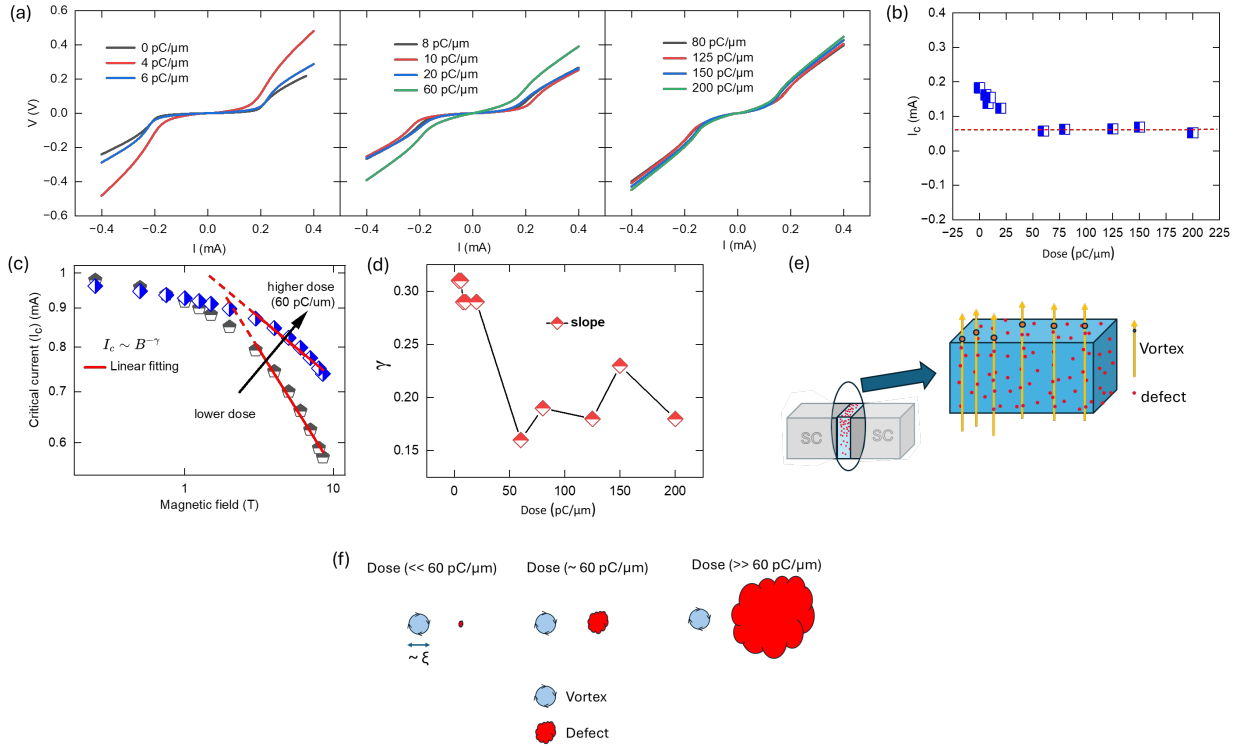

**Figure 5.** Current-Voltage Characteristics for Different Doses. (a) $I$-$V$ characteristics for devices irradiated with different doses of $Si^{++}$ ion beam. A gradual evolution in the $I$-$V$ behavior is observed as the dose increases from 0 to 200 pC/μm. (b) The extracted critical current ($I_c$) from (a), is plotted as a function of dose, revealing a decreasing trend of $I_c$ with increasing dose. Eventually, $I_c$ saturates beyond a certain dose, indicating a limit to the suppression of superconductivity. (c) The log-log plot of critical current ($I_c$) versus magnetic field ($B$) is shown for two devices fabricated with different exposure doses: a low dose (4 pC/μm) and a higher dose (60 pC/μm). The scaling exponent ($\gamma$) is extracted from the $I_c$ vs $B^{-\gamma}$ relationship using linear fits (shown in red line). It is evident that the critical current decreases more rapidly with increasing magnetic field in the low-dose devices compared to the high-dose ones. (d) The scaling factor is now extracted for all the different devices with different doses and plotted against the dose values. Notably, despite the reduction in $I_c$, the superconducting state becomes more resistant to magnetic field-induced degradation due to enhanced vortex pinning from ion irradiation. This effect is reflected in the initial decrease of $\gamma$ with dose. The eventual saturation of $\gamma$ at higher doses suggests that excessive irradiation does not further enhance pinning. (e) schematics showing the vortex pinning to the defects created by ion bombardment. The symbols for the defects and the vortices are also labelled in the figure. (f) The schematic illustrates various scenarios based on the size differences between defects and vortices. In the central scenario, where the sizes are closely matched, the defects are most effective at pinning the vortices. In contrast, the other two cases, where there is a significant size disparity, show less effective vortex pinning.

When a magnetic field penetrates a superconductor, localized non-superconducting regions (vortices) form on the scale of the coherence length. These vortices are accompanied by circulating supercurrents and experience a Lorentz force that drives them to move [49]; being normal regions, their motion causes dissipation and further weakens superconductivity. Vortex motion can be mitigated by defects that act as pinning sites [50]. In our devices, focused $Si^{++}$ ion bombardment introduces controlled defects within the FTS/BT heterostructure, serving as vortex pinning centers [36,51]. This is illustrated in Figure 5c and 5d. Figure 5c shows a log-log plot of $I_c$ versus B for low dose (4 pC/$\mu$m) and higher dose (60 pC/$\mu$m) devices; linear fits (red lines) yield the power law exponent $\gamma$ in $I_c \sim B^{-\gamma}$. The higher dose device exhibits a slower decay of $I_c$ with B, implying stronger pinning. Figure 5d plots $\gamma$ versus dose for multiple devices. Initially, $\gamma$ decreases with increasing dose, indicating enhanced vortex pinning and thus more robust superconductivity. Beyond a critical defect density (~60 pC/$\mu$m), however, $\gamma$ saturates, suggesting that further defects do not significantly increase pinning efficacy. A schematic of this vortex pinning mechanism is shown in Figure 5e, where vortices (orange circles with arrows) are anchored to ion-induced defect sites (red dots) in the heterostructure. One striking feature of non-monotonic relationship with dose is evident for all the parameters studied here, such as Tc, Ic and $\gamma$ (Figure 3b, 5b and 5d). All three parameters start changing rapidly at lower dose and this continues to occur till the dose of ~ 60 pc/um and then these parameters saturate beyond this point. Superconducting performance in irradiated Fe(Te,Se)/Bi$_2$Te$_3$ thin films is governed by two defect-controlled factors: the number of defects and their size. Each defect can act simultaneously as a scattering center, which disrupts Cooper-pair coherence and suppresses superconductivity, and as a pinning center, which immobilizes vortices and curbs flux creep. The balance between these antagonistic roles underlies the complex dose dependence of the critical current density Ic and transition temperature Tc.

Microscopy of Fe(Te,Se) single crystals bombarded by high-energy $Au^{2+}$ ions has mapped this defect landscape in detail [36]. Assuming a comparable defect profile for our thin-film geometry, despite our use of focused, lower-energy $Si^{2+}$ beams, we infer that low doses chiefly introduce point-like defects. Because these defects are similar in size to the coherence length ($\xi \approx 2$ nm), they reinforce vortex pinning without introducing excessive scattering. Accordingly, in a magnetic field the films tolerate higher vortical loads, whereas in zero field the omnipresent scattering lowers both Ic and Tc.

As the dose rises, neighboring point defects merge into clusters. The overall number of scattering centers increases only modestly, leading to the observed saturation of Ic and Tc; however, the size of the clusters soon exceeds $\xi$, diminishing their pinning efficacy. This crossover appears in the $\gamma$–dose curve at $\approx 60$ pC $\mu m^{-1}$ (Figure 1d). Beyond this threshold the enlarged clusters no longer match the vortex core size, and pinning efficiency falls.

Vortex–defect matching is schematized in Figure 5f. Very small (left) or very large (right) defects fail to pin vortices effectively, whereas an intermediate size (center) provides optimal pinning. Transport measurements corroborate this picture: initial doses create angstrom-scale vacancies and interstitials; increasing doses promote their coalescence into 2–3 nm clusters around 60 pC $\mu m^{-1}$, boosting pinning; still higher doses generate amorphous regions that undermine vortex immobilization. STM imaging of ion-irradiated Fe(Te,Se) confirms this evolution of defect morphology [36].

## 3. Conclusions:

In this work, we have investigated the effect of a focused $Si^{++}$ ion beam nanosculpting on a microbridge fabricated from a thin film of a topological superconductor candidate $FeTe_{0.75}Se_{0.25}/Bi_2Te_3$, enabling us to fabricate a lateral junction of superconductor-irradiated nanoregion-superconductor. The irradiated regions (starting from ion dose of ~150 pC/$\mu$m) were characterized by scanning electron microscopy (SEM), atomic force microscopy (AFM), and Kelvin probe force microscopy (KPFM), followed by detailed electrical transport measurements including resistance vs. temperature and dose-dependent $I$-$V$ studies. Our results show that both the $T_c$ and the $I_c$ decrease with increasing ion dose, indicating a progressive weakening of superconductivity due to irradiation. By adjusting the ion beam dose, we effectively controlled the defect density, which in turn enhanced vortex pinning. Consequently, $I_c$ demonstrated a slower decline with increasing magnetic field for higher irradiation doses. This outcome underscores the potential of defect engineering in creating vortex pinning sites, which have been theoretically predicted to host the long-sought Majorana modes in a topological superconductor [52-54].

The changes observed in the critical temperature (Tc) and critical current (Ic) provide evidence of the successful formation of ion-irradiated localized insulating junction by nano-sculpting. Significant deviations in these parameters would have indicated substantial damage, rather than the establishment of a weak link. The strength of this weak link will ultimately determine whether the junction can exhibit the Fraunhofer pattern, which arises from the diffraction of the Cooper pair wave function through the non-superconducting layer. We believe that the absence of the Fraunhofer pattern is likely due to the increased width of the middle layer which becomes mismatched with the superconducting coherence length and thus precludes the observable diffraction of Cooper pairs. Further investigation is required to observe Fraunhofer patterns or the creation of other device structures such as superconducting quantum interference devices (SQUIDs).

Notably, this technique has significant scalability potential and this is its first application utilizing exotic chalcogenide-based superconductors. We believe that our approach offers exciting opportunities for future research in this area toward realizing fault-tolerant qubits for quantum computing applications as well as a number of other use inspired applications based on chalcogenide materials.

## 4. Experimental Section

*Thin film synthesis:* The thin film heterostructure, $FeTe_{0.75}Se_{0.25}/Bi_2Te_3$ was grown on $Al_2O_3(0001)$ substrate by molecular beam epitaxy (MBE) in a vacuum chamber with a base pressure of $10^{-10}$ Torr. The growth details can be found in [23] and the supplementary information therein.

*Device Fabrication:* UV photolithography was employed to define the initial bar pattern for each device. After exposure and development of the bar pattern, reactive ion etching (RIE) using a gas mixture of $BCl_3$ and $Cl_2$ in a 1:3 ratio was used at 800 W, which yielded 16 devices, each featuring four probes and a 5 μm wide channel for subsequent ion exposure.

*Focused ion beam irradiation:* The devices were mounted inside a Raith Velion focused ion beam system and irradiated the devices using a 35 kV $Si^{++}$ ion beam (~ 6 nm beam diameter) so that landing energies of ions were 70 keV. Dose was varied 0 pC/μm to 200 pC/μm by adjusting pattern time. During ion irradiation, the central regions of the devices were shielded from exposure to prevent unintended ion-induced damage. The instrument was equipped with *in situ* micromanipulators for electrical contact, which were connected to a Keithley 2400 that was used to apply the constant voltage bias and to measure the current during ion irradiation.

*AFM and KPFM:* Atomic Force Microscopy (AFM) and Kelvin Probe Force Microscopy (KPFM) measurements were performed using an Asylum Research Cypher AFM operating in lift mode. The sample device was wire-bonded to breakout cables and grounded through the electrical connections to ensure stable potential referencing during scanning. For KPFM, Pt-coated Multi75G probes (nominal resonance frequency ~75 kHz, spring constant ~3 N/m, and tip radius <25 nm) were employed to enable high resolution surface potential mapping. An AC voltage of 1 V was applied at the cantilever's resonance frequency to excite the electrostatic force gradient, while lift mode scanning was conducted approximately 30 nm above the sample surface to decouple topographic and electrostatic signals. This configuration enabled sensitive and localized work function measurements with minimized crosstalk from surface morphology.

*Ion / matter interaction simulations via SRIM:* To calculate the ion trajectories and energy loss profiles, SRIM was used with the with damage calculations set to "Monolayer Collisions". $Si^{++}$ ions were approximated as $Si^+$ ions at 70 keV since the actual experimental condition is $Si^{++}$ ions in a 35 kV accelerating voltage. Trajectories originated from a point source at normal incidence to the surface. The electronic densities of each layer set were set to 6.24 g/cm$^3$ for Te, 6.40 g/cm$^3$ for $FeTe_{0.75}Se_{0.25}$, 7.84g /cm$^3$ for $Bi_2Te_3$, and 3.98 g/cm$^3$ for $Al_2O_3$. The thickness for each layer was the same as reported in Figure 1(a). Simulations of 5,000 ions were used to ensure sufficient statistics.

*Ex situ electrical transport measurements:* All the devices were measured in a Quantum Design Physical Properties Measurement System (PPMS), with a base temperature of 2 K and a maximum magnetic field of 9 T. We used indium wires to make the electrical contacts from the sample to the puck.

## Supporting Information

Supporting Information is available from the Wiley Online Library or from the author.

## Acknowledgements

This material is based upon work supported by the U.S. Department of Energy, Office of Science, National Quantum Information Science Research Centers, Quantum Science Center. The photolithography, AFM, KPFM, and FIB were performed under a user project at the Center for Nanophase Materials Sciences, which is a Department of Energy, Office of Science User Facility at Oak Ridge National Laboratory. S.G., L.C., S.J, S.J.R., and S.T.R, are partially supported by the Center for Nanophase Materials Sciences.

## Conflict of Interest

The authors declare no conflict of interest.

## Data Availability Statement

The data that support the findings of this study are available from the corresponding author upon reasonable request.

Received: ((will be filled in by the editorial staff))

Revised: ((will be filled in by the editorial staff))

Published online: ((will be filled in by the editorial staff))

## References

[1] Kamihara Y, Watanabe T, Hirano M, Hosono H. Iron-based layered superconductor La [O1-x F x] FeAs (x= 0.05- 0.12) with T c= 26 K. *Journal of the American Chemical Society* 2008; 130(11): 3296–3297.

[2] Fernandes RM, Coldea AI, Ding H, Fisher IR, Hirschfeld P, Kotliar G. Iron pnictides and chalcogenides: a new paradigm for superconductivity. *Nature* 2022; 601(7891): 35–44.

[3] Biswal G, Mohanta K. A recent review on iron-based superconductor. *Materials Today: Proceedings* 2021; 35: 207–215.

[4] Hosono H, Kuroki K. Iron-based superconductors: Current status of materials and pairing mechanism. *Physica C: Super- conductivity and its Applications* 2015; 514: 399–422.

[5] Ge JF, Liu ZL, Liu C, et al. Superconductivity above 100 K in single-layer FeSe films on doped SrTiO 3. *Nature materials* 2015; 14(3): 285–289.

[6] Wang F, Lee DH. The electron-pairing mechanism of iron-based superconductors. *Science* 2011; 332(6026): 200–204.

[7] Hirschfeld PJ. Using gap symmetry and structure to reveal the pairing mechanism in Fe-based superconductors. *Comptes Rendus Physique* 2016; 17(1-2): 197–231.

[8] Wang QY, Li Z, Zhang WH, et al. Interface-Induced High-Temperature Superconductivity in Single Unit-Cell FeSe Films on SrTiO 3. *Chinese Physics Letters* 2012; 29(3).

[9] Lee JJ, Schmitt FT, Moore RG, et al. Interfacial mode coupling as the origin of the enhancement of Tc in FeSe films on SrTiO3. *Nature* 2014; 515(7526): 245-248.

[10] Zhang P, Yaji K, Hashimoto T, et al. Observation of topological superconductivity on the surface of an iron-based super- conductor. *Science* 2018; 360(6385): 182–186.

[11] Wu X, Chung SB, Liu C, Kim EA. Topological orders competing for the Dirac surface state in FeSeTe surfaces. *Physical Review Research* 2021; 3(1): 013066.

[12] Wang Z, Zhang P, Xu G, et al. Topological nature of the FeSe 0.5 Te 0.5 superconductor. *Physical Review B* 2015; 92(11): 115119.

[13] Xu G, Lian B, Tang P, Qi XL, Zhang SC. Topological superconductivity on the surface of Fe-based superconductors. *Physical review letters* 2016; 117(4): 047001.

[14] Kitaev AY. Fault-tolerant quantum computation by anyons. *Annals of Physics* 2003; 303(1): 2-30.

[15] Li Y, Zaki N, Garlea VO, et al. Electronic properties of the bulk and surface states of Fe(1+y)Te(1-x)Se(x). *Nature Materials* 2021; 20(9): 1221-1227.

[16] Fu L, Kane CL. Superconducting proximity effect and majorana fermions at the surface of a topological insulator. *Physical Review Letters* 2008; 100(9): 096407.

[17] He QL, Liu H, He M, et al. Two-dimensional superconductivity at the interface of a Bi2Te3/FeTe heterostructure. *Nature communications* 2014; 5(1): 4247.

[18] Mallick D, Yi HT, Yuan X, Oh S. Ubiquity of Rotational Symmetry Breaking in Superconducting Films, From Fe (Te, Se)/Bi2Te3 to Nb, and the Effect of Measurement Geometry. *Advanced Science* 2025: 2504430.

[19] Chen M, Chen X, Yang H, Du Z, Wen HH. Superconductivity with twofold symmetry in Bi2Te3/FeTe0. 55Se0. 45 heterostructures. *Science Advances* 2018; 4(6): eaat1084.

[20] Guo B, Shi KG, Qin HL, et al. Evidence for topological superconductivity: Topological edge states in Bi2Te3/FeTe heterostructure. *Chinese Physics B* 2020; 29(9): 097403.

[21] Yao X, Brahlek M, Yi HT, et al. Hybrid symmetry epitaxy of the superconducting Fe (Te, Se) film on a topological insulator. *Nano letters* 2021; 21(15): 6518–6524.

[22] Moore RG, Lu Q, Jeon H, et al. Monolayer Superconductivity and Tunable Topological Electronic Structure at the Fe(Te,Se)/Bi(2) Te(3) Interface. *Advanced Materials* 2023; 35(22): e2210940.

[23] Chen AH, Lu Q, Hershkovitz E, et al. Interfacially Enhanced Superconductivity in Fe (Te, Se)/Bi4Te3 Heterostructures. *Advanced Materials* 2024; 36(31): 2401809.

[24] Kobayashi Y, Shiogai J, Nojima T, Matsuno J. A scaling relation of vortex-induced rectification effects in a superconducting thin-film heterostructure. *Communications Physics* 2025; 8(1): 1–6.

[25] Blatter G, Feigel'man MV, Geshkenbein VB, Larkin AI, Vinokur VM. Vortices in high-temperature superconductors. *Reviews of modern physics* 1994; 66(4): 1125.

[26] Civale L. Vortex pinning and creep in high-temperature superconductors with columnar defects. *Superconductor Science and Technology* 1997; 10(7A): A11.

[27] Miura M, Tsuchiya G, Harada T, et al. Thermodynamic approach for enhancing superconducting critical current performance. *NPG Asia Materials* 2022; 14(1): 85.

[28] Puig T, Gutierrez J, Obradors X. Impact of high growth rates on the microstructure and vortex pinning of high-temperature superconducting coated conductors. *Nature Reviews Physics* 2024; 6(2): 132–148.

[29] Miura M, Eley S, Iida K, et al. Quadrupling the depairing current density in the iron-based superconductor SmFeAsO1–x H x. *Nature Materials* 2024; 23(10): 1370–1378.

[30] Sun Y, Pyon S, Tamegai T, et al. Enhancement of critical current density and mechanism of vortex pinning in H+-irradiated FeSe single crystal. *Applied Physics Express* 2015; 8(11): 113102.

[31] Fang L, Jia Y, Mishra V, et al. Huge critical current density and tailored superconducting anisotropy in SmFeAsO0. 8F0. 15 by low-density columnar-defect incorporation. *Nature Communications* 2013; 4(1): 2655.

[32] Nakajima Y, Tsuchiya Y, Taen T, Tamegai T, Okayasu S, Sasase M. Enhancement of critical current density in Co-doped BaFe 2 As 2 with columnar defects introduced by heavy-ion irradiation. *Physical Review B—Condensed Matter and Mate- rials Physics* 2009; 80(1): 012510.

[33] Kihlstrom K, Fang L, Jia Y, et al. High-field critical current enhancement by irradiation induced correlated and random defects in (Ba0. 6K0. 4) Fe2As2. *Applied Physics Letters* 2013; 103(20).

[34] Tamegai T, Taen T, Yagyuda H, et al. Effects of particle irradiations on vortex states in iron-based superconductors. *Superconductor Science and Technology* 2012; 25(8): 084008.

[35] Strickland N, Talantsev E, Long N, et al. Flux pinning by discontinuous columnar defects in 74 MeV Ag-irradiated YBa2Cu3O7 coated conductors. *Physica C: Superconductivity* 2009; 469(23-24): 2060–2067.

[36] Massee F, Sprau PO, Wang YL, et al. Imaging atomic-scale effects of high-energy ion irradiation on superconductivity and vortex pinning in Fe (Se, Te). *Science Advances* 2015; 1(4): e1500033.

[37] Ozaki T, Wu L, Gu G, Li Q. Ion irradiation of iron chalcogenide superconducting films. *Superconductor Science and Technology* 2020; 33(9): 094008.

[38] Zheng Y, Li S, Ding Z, Xiong K, Feng J, Yang H. Fabrication of Al/AlOx/Al junctions with high uniformity and stability on sapphire substrates. *Scientific Reports* 2023; 13(1): 11874.

[39] Ruhtinas A, Maasilta IJ. Highly tunable NbTiN Josephson junctions fabricated with focused helium ion beam. *arXiv preprint arXiv:2303.17348* 2023.

[40] Cybart SA, Cho E, Wong T, et al. Nano Josephson superconducting tunnel junctions in YBa2Cu3O7–δ directly patterned with a focused helium ion beam. *Nature nanotechnology* 2015; 10(7): 598–602.

[41] Müller B, Karrer M, Limberger F, et al. Josephson junctions and squids created by focused helium-ion-beam irradiation of y ba 2 cu 3 o 7. *Physical Review Applied* 2019; 11(4): 044082.

[42] Li H, Cai H, Forman J, et al. Transport properties of NbN thin films patterned with a focused helium ion beam. *IEEE Transactions on Applied Superconductivity* 2023; 33(5): 1–4.

[43] Fornieri A, Whiticar AM, Setiawan F, et al. Evidence of topological superconductivity in planar Josephson junctions. *Nature* 2019; 569(7754): 89–92.

[44] Pientka F, Keselman A, Berg E, Yacoby A, Stern A, Halperin BI. Topological superconductivity in a planar Josephson junction. *Physical Review X* 2017; 7(2): 021032.

[45] Stoller RE, Toloczko MB, Was GS, Certain AG, Dwaraknath S, Garner FA. On the use of SRIM for computing radiation damage exposure. *Nuclear instruments and methods in physics research section B: beam interactions with materials and atoms* 2013; 310: 75–80.

[46] Stanford MG, Pudasaini PR, Belianinov A, et al. Focused helium-ion beam irradiation effects on electrical transport properties of few-layer WSe2: enabling nanoscale direct write homo-junctions. *Scientific reports* 2016; 6(1): 27276.

[47] Jung SG, Son SK, Pham D, et al. Influence of carbon-ion irradiation on the superconducting critical properties of MgB2 thin films. *Superconductor Science and Technology* 2019; 32(2): 025006.

[48] Sueyoshi T, Nishimura T, Fujiyoshi T, Mitsugi F, Ikegami T, Ishikawa N. High flux pinning efficiency by columnar defects dispersed in three directions in YBCO thin films. *Superconductor Science and Technology* 2016; 29(10): 105006.

[49] Anderson PW, Kim Y. Hard superconductivity: theory of the motion of Abrikosov flux lines. *Reviews of modern physics* 1964; 36(1): 39.
[50] Kramer EJ. Scaling laws for flux pinning in hard superconductors. *Journal of applied physics* 1973; 44(3): 1360–1370.
[51] Ishida S, Iyo A, Ogino H, et al. Unique defect structure and advantageous vortex pinning properties in superconducting CaKFe4As4. *npj Quantum Materials* 2019; 4(1): 27.
[52] Liu Q, Chen C, Zhang T, et al. Robust and clean Majorana zero mode in the vortex core of high-temperature superconductor (Li 0.84 Fe 0.16) OHFeSe. *Physical Review X* 2018; 8(4): 041056.
[53] Chiu CK, Machida T, Huang Y, Hanaguri T, Zhang FC. Scalable Majorana vortex modes in iron-based superconductors. *Science Advances* 2020; 6(9): eaay0443.
[54] Fan P, Yang F, Qian G, et al. Observation of magnetic adatom-induced Majorana vortex and its hybridization with field- induced Majorana vortex in an iron-based superconductor. *Nature Communications* 2021; 12(1): 1348.
[55] T., Wu, L., Zhang, C. et al. A route for a strong increase of critical current in nanostrained iron-based superconductors. *Nature Communications* 2016; **7**, 13036.

**Nanosculpting lateral weak link junctions in superconducting Fe(Te,Se)/Bi$_2$Te$_3$ with focused Si++ ions and implications on vortex pinning**